\documentclass[a4paper,aps,prd,12pt,preprintnumbers,
onecolumn,superscriptaddress,nofootinbib,
amsmath,amssymb]{revtex4-1}
\usepackage{setspace}
\usepackage{graphicx,multirow,cmap,ulem}
\usepackage[utf8]{inputenc}
\usepackage[T1]{fontenc}

\def\re#1{Re(#1)}
\def\im#1{Im(#1)}

\def\Order#1{{\cal O}\left(#1\right)}

\begin{document}

\doublespacing
\title{Scattering of a scalar field in the four-dimensional quasi-topological gravity}
\author{Alexey Dubinsky}
\email{\texttt{dubinsky@ukr.net}}
\affiliation{University of Seville, Seville, Spain}
\begin{abstract}
We study grey-body factors for a massless scalar field in the spacetime of regular black holes arising in four–dimensional non-polynomial quasi-topological gravity. We consider two representative metrics that capture the typical features of regular geometries. Using the WKB method, we compute the transmission probabilities and analyze their dependence on the regularization parameter. The grey-body factors are found to deviate only slightly from the Schwarzschild case, indicating that the scattering properties are largely insensitive to near-horizon regularization of the geometry. The correspondence between quasinormal modes and grey-body factors is shown to be sufficiently accurate for higher multipole numbers. 
\end{abstract}
\maketitle

\section{Introduction}

Understanding the behavior of gravity in the strong-field regime remains one of the central problems of modern theoretical physics. In particular, the classical description of black holes predicts the formation of spacetime regions where curvature invariants diverge and the classical evolution of geodesics breaks down. These pathologies are generally interpreted as an indication that classical general relativity ceases to be applicable at sufficiently high curvatures and that additional physical ingredients must intervene to regulate the geometry. Consequently, considerable attention has been devoted to the search for black-hole solutions in which the interior structure is modified so that curvature singularities disappear while the exterior stays close to the Schwarzschild or Kerr geometry.

 One promising resolution is provided by {\it regular black holes}, in which the central singularity is replaced by a smooth geometry with finite curvature invariants. In many such models, the near–origin region acquires an effective de Sitter–like behavior, preventing the divergence of curvature scalars and ensuring geodesic completeness of the spacetime.

Historically, most regular black-hole solutions were constructed phenomenologically. This was achieved either by coupling Einstein gravity to exotic matter sources or by using effective models inspired by quantum gravity \cite{Bronnikov:2000vy,Konoplya:2025ect,Dymnikova:1992ux,Ansoldi:2008jw,
Bonanno:2000ep,AyonBeato:1998ub,Bronnikov:2024izh,Hayward:2005gi,
Spina:2025wxb,Bardeen:1968,Dymnikova:2015yma,Bronnikov:2005gm}. 
While such models offer valuable insight into singularity resolution, their matter content is often introduced ad hoc and does not necessarily follow from fundamental principles or a well-defined theory of gravity. An alternative and conceptually appealing possibility is that regular black holes emerge as vacuum solutions of modified gravity theories containing higher–curvature corrections. Such corrections are expected to appear naturally in effective descriptions of gravity, for instance as quantum or string-inspired contributions that become important in the strong-curvature regime.

Another motivation for studying regular geometries comes from the perspective of black-hole perturbation theory and gravitational-wave astronomy. The ringdown phase of a perturbed black hole encodes information about the near-horizon geometry through its quasinormal spectrum. Consequently, deviations from the Schwarzschild structure near the center or the horizon may leave observable imprints on the oscillation frequencies and damping rates of the modes. Regular black holes therefore provide an interesting laboratory for investigating how modifications of the strong-field geometry affect dynamical observables.

Recently, it has been shown that regular black holes may arise naturally in four-dimensional non-polynomial quasi-topological gravity \cite{Borissova:2026wmn}. In this framework, static and spherically symmetric vacuum solutions can be obtained in closed form, while the field equations reduce to a remarkably simple algebraic relation for the metric function. The resulting geometries are asymptotically flat and free of curvature singularities, reproducing well-known regular metrics such as the Hayward and Dymnikova solutions without introducing additional matter fields. Within this class of solutions, one can construct a generic metric profile depending on a small number of parameters controlling the deviation from the Schwarzschild geometry. 

These parameters effectively determine the structure of the near-core region and the extent to which the geometry differs from the classical vacuum solution. In particular, they may regulate the transition between the Schwarzschild-like exterior and the regular interior, thereby influencing both the horizon structure and the behavior of perturbations propagating in the spacetime. The study of wave dynamics in such backgrounds, therefore, provides an important tool for understanding the physical consequences of the regularization mechanism.

Understanding how such geometrically regular black holes interact with propagating fields is essential for determining their observational signatures. In addition to the quasinormal spectrum that governs the ringdown stage of perturbations \cite{Konoplya:2026gim}, another important characteristic of black-hole spacetimes is the frequency-dependent transmission probability of waves through the effective potential barrier surrounding the event horizon \cite{Page:1976df}. These transmission probabilities are commonly known as {\it grey-body factors}. They describe the modification of the purely thermal Hawking radiation spectrum due to scattering in the curved spacetime geometry and therefore play a key role in determining the observable energy emission rates of black holes \cite{Page:1976ki,Cvetic:1997xv,Gubser:1996zp,Klebanov:1997cx,Kanti:2014vsa}.

Grey-body factors have been extensively studied for a wide variety of black-hole geometries, including asymptotically flat, de Sitter, and anti–de Sitter spacetimes, as well as for numerous modified gravity scenarios (see, for recent examples, \cite{Kanti:2002ge,Dubinsky:2024vbn,
Lutfuoglu:2025ljm,Bonanno:2025dry,Tang:2025mkk,Kanti:2017ubd,
Dubinsky:2025nxv,Lutfuoglu:2025blw,Skvortsova:2024msa,
Malik:2025erb,Duffy:2005ns,Han:2025cal,Lutfuoglu:2025hjy,Kanti:2002nr,
Fernando:2016ksb,Arbelaez:2026eaz,Pappas:2016ovo,Bolokhov:2025lnt,Harris:2005jx,
Dubinsky:2024nzo,Lutfuoglu:2025ldc,Malik:2025dxn,Gohain:2024aod}). In asymptotically flat backgrounds, the grey-body factor corresponds to the absorption probability of an incoming wave scattered by the effective potential barrier, while the complementary reflection coefficient describes the portion of radiation returning to spatial infinity. These quantities encode detailed information about the structure of the effective potential and, therefore, provide a useful probe of deviations from the Schwarzschild geometry.

Quasi-topological gravity offers an elegant framework for constructing regular black-hole spacetimes \cite{Bueno:2024dgm,Bueno:2024eig,Bueno:2025tli}.
In the present work, we investigate grey-body factors for a test scalar field propagating in the spacetime of regular black holes arising in four-dimensional non-polynomial quasi-topological gravity \cite{Bueno:2025zaj,Borissova:2026wmn}. We focus on two representative metric functions found in \cite{Borissova:2026wmn}, which illustrate the typical behavior of such geometries. 

Our goal is to determine how the near-horizon modifications responsible for regularizing the geometry influence the scattering properties of fields and the corresponding absorption probabilities. Since grey-body factors are determined by the same effective potentials that govern quasinormal oscillations, they provide complementary information about the interaction between waves and the gravitational potential barrier. In particular, deviations of the effective potential from the Schwarzschild case may lead to characteristic shifts in the transmission spectra and energy emission rates.

The paper is organized as follows. In the next section, we briefly introduce the two specific metrics considered in this work. We then derive the perturbation equations for scalar fields and formulate the scattering problem relevant for the computation of grey-body factors. Subsequently, we evaluate the transmission probabilities and absorption cross sections using the WKB approach and analyze their dependence on the parameters controlling the deviation from the Schwarzschild geometry. We also test the correspondence between grey-body factors and quasinormal modes. Finally, we summarize our results and discuss their physical implications.

\section{Regular black hole metrics in quasi-topological gravity}

We now consider the specific class of regular black hole metrics arising from four-dimensional non-polynomial quasi-topological gravity. In this framework, the field equations for a static spherically symmetric ansatz reduce to a single algebraic equation for the metric function, allowing solutions to be found in closed form.

We begin with the general static spherically symmetric line element
\begin{equation}
ds^{2}=-f(r)dt^{2}+\frac{dr^{2}}{f(r)}+r^{2}d\Omega^{2}.
\end{equation}

For a broad class of models within this theory, the metric function can be expressed in the following generic form \cite{Tsuda:2026xjc,Konoplya:2026gim}
\begin{equation}
f(r)=1-
\frac{2Mr^{\mu-1}}
{\left(r^{\nu}+\alpha^{\nu/3}(2M)^{\nu/3}\right)^{\mu/\nu}},
\end{equation}
where $M$ is an integration constant identified with the asymptotic mass and $\alpha$ (with dimensions of length squared) is a parameter that controls deviations from the Schwarzschild geometry. It is convenient to denote $\alpha=l^{2}$ and measure all dimensional quantities in units of the black hole mass $M$.

In the present work, we consider two particular models, one of which arises from the above general solution for specific values of the integers $\mu$ and $\nu$. The first model suggested in \cite{Borissova:2026wmn} corresponds to
\begin{equation}
f(r)=1-
\frac{2Mr^{2}}{\sqrt{4l^{4}M^{2}+r^{6}}},
\end{equation}
which represents a regular black hole solution obtained within the quasi–topological gravity framework. The parameter $l$ controls the strength of regularization and determines the deviation of the metric from the Schwarzschild spacetime.

The second model we consider falls outside the general class and  is given by the following metric function \cite{Borissova:2026wmn},
\begin{equation}
f(r)=1-\frac{r^2 \left(1-e^{-\frac{l^2 M}{r^3}}\right)}{l^2}.
\end{equation}

Both metrics are asymptotically flat and reduce to the Schwarzschild solution in the limit $l\to0$. At the same time, the geometry remains regular at $r=0$, where curvature invariants remain finite. The deviation from the Schwarzschild spacetime is mainly localized in the near–horizon region, where the effective potential governing wave propagation is modified.

In the following sections, we analyze the propagation and scattering of a test massless scalar field in these two regular black hole backgrounds and compute the corresponding grey–body factors.

\begin{figure}
\resizebox{\linewidth}{!}{\includegraphics{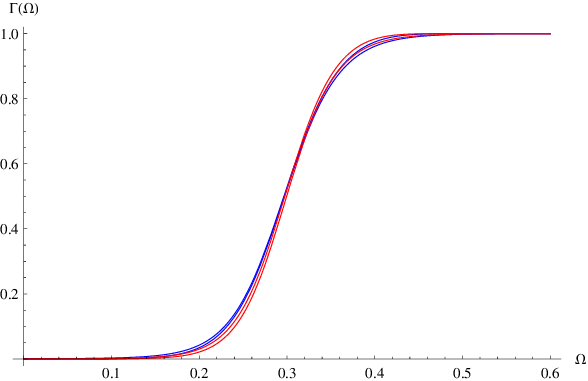}\includegraphics{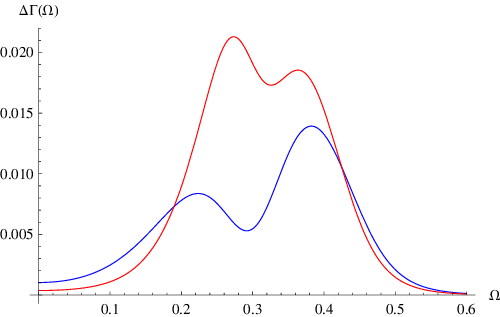}}
\caption{Grey-body factors for the black-hole model I: $\ell=1$, $l=0.1$ (blue) and $l=1.24$ (red) $M=1$.}\label{fig:GBFs1}
\end{figure}

\begin{figure}
\resizebox{\linewidth}{!}{\includegraphics{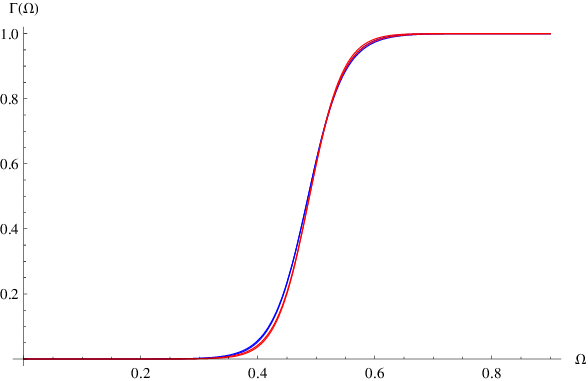}\includegraphics{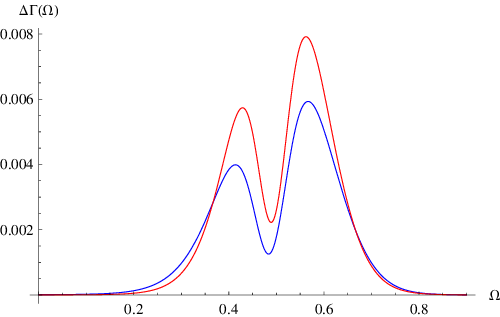}}
\caption{Grey-body factors for the black-hole model I: $\ell=2$, $l=0.1$ (blue) and $l=1.24$ (red) $M=1$.}\label{fig:GBFs2}
\end{figure}

\begin{figure}
\resizebox{\linewidth}{!}{\includegraphics{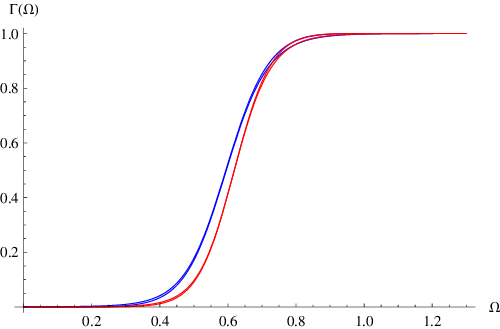}\includegraphics{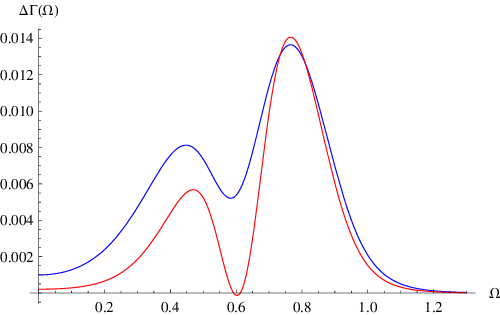}}
\caption{Grey-body factors for the black-hole model II: $\ell=1$, $l=0.1$ (blue) and $l=0.5$ (red) $M=1$.}\label{fig:GBFs3}
\end{figure}

\begin{figure}
\resizebox{\linewidth}{!}{\includegraphics{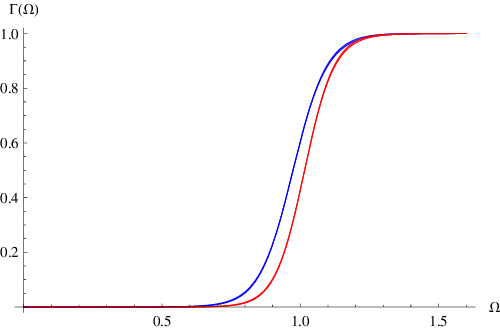}\includegraphics{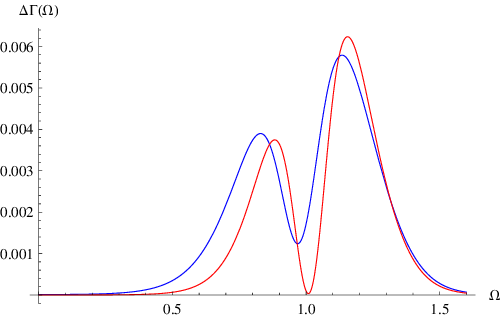}}
\caption{Grey-body factorsfor the black-hole  model II: $\ell=2$, $l=0.1$ (blue) and $l=0.5$ (red) $M=1$.}\label{fig:GBFs4}
\end{figure}

\section{The scattering problem}

We consider the propagation of a test massless scalar field in the background of a static, spherically symmetric black hole spacetime. The field obeys the covariant Klein--Gordon equation
\begin{equation}
\frac{1}{\sqrt{-g}}\partial_{\mu}\left(\sqrt{-g}\,g^{\mu\nu}\partial_{\nu}\Phi\right)=0 .
\end{equation}
For a general metric of the form
\begin{equation}
ds^{2}=-f(r)dt^{2}+\frac{dr^{2}}{f(r)}+r^{2}d\Omega^{2},
\end{equation}
we decompose the scalar field as
\begin{equation}
\Phi(t,r,\theta,\phi)=e^{-i\omega t}\frac{\Psi(r)}{r}Y_{\ell m}(\theta,\phi),
\end{equation}
where $Y_{\ell m}$ are spherical harmonics. Substituting this ansatz into the wave equation yields a Schr\"odinger--type radial equation
\begin{equation}
\frac{d^{2}\Psi}{dr_*^{2}}+\left(\omega^{2}-V(r)\right)\Psi=0.
\end{equation}
Here, the tortoise coordinate $r_*$ is defined by
\begin{equation}
\frac{dr_*}{dr}=\frac{1}{f(r)} ,
\end{equation}
which maps the event horizon $r=r_h$ to $r_*\to -\infty$ and spatial infinity to $r_*\to +\infty$. 
For a massless scalar field, the effective potential is
\begin{equation}
V(r)=f(r)\left(\frac{\ell(\ell+1)}{r^{2}}+\frac{f'(r)}{r}\right).
\end{equation}

In asymptotically flat black hole spacetimes, this effective potential vanishes at both boundaries,
\begin{equation}
V(r)\rightarrow 0 , \qquad r_*\rightarrow \pm\infty ,
\end{equation}
forming a potential barrier outside the event horizon.
The scattering of waves by this barrier determines the grey-body factors.

In the scattering problem one considers a wave incident from spatial infinity toward the black hole, and the asymptotic solution may be expressed as
\begin{equation}
\Psi \sim 
\begin{cases}
T\,e^{-i\omega r_*}, & r_*\rightarrow -\infty,\\
e^{-i\omega r_*}+R\,e^{i\omega r_*}, & r_*\rightarrow +\infty ,
\end{cases}
\end{equation}
where $T$ is the transmission amplitude (the part of the wave absorbed by the black hole) and $R$ is the reflection coefficient. Conservation of the Wronskian implies
\begin{equation}
|R|^{2}+|T|^{2}=1 .
\end{equation}

The grey-body factor, or absorption probability, for a given multipole number $\ell$ is defined as the transmission probability
\begin{equation}
\Gamma_{\ell}(\omega)=|T|^{2}.
\end{equation}
Physically, it represents the probability that a wave originating from infinity penetrates the potential barrier and reaches the event horizon. By time-reversal symmetry of the S-matrix, it also equals the probability that a wave emanating from the horizon reaches a distant observer. These factors modify the Hawking radiation spectrum and determine the black hole's absorption cross-section.

Within the WKB approximation, the transmission probability through a potential barrier can be expressed in terms of the properties of the effective potential near its maximum. For a potential with a single peak at $r=r_0$ the WKB formula gives \cite{Schutz:1985km,Iyer:1986np}
\begin{equation}
\Gamma_{\ell}(\omega)=\frac{1}{1+\exp(2\pi iK)},
\end{equation}
where the quantity $K$ follows from the WKB expansion \cite{Schutz:1985km,Iyer:1986np,Matyjasek:2017psv,Konoplya:2003ii}
\begin{equation}
K=i\frac{\omega^{2}-V_0}{\sqrt{-2V_0''}}+\Lambda_2+\Lambda_3+\cdots .
\end{equation}
Here, $V_0$ denotes the value of the effective potential at its maximum and $V_0''$ is the second derivative of the potential with respect to the tortoise coordinate  at that point. The terms $\Lambda_i$ represent higher-order WKB corrections that depend on progressively higher derivatives of the potential at the maximum. 
Explicit expressions for these corrections are known up to high orders and allow one to compute grey-body factors with good accuracy for a wide range of frequencies.

The WKB method provides a particularly efficient approach for computing grey-body factors because it requires only local knowledge of the effective potential near its peak.  For low multipole numbers, the method remains reliable, provided sufficiently high orders of the WKB expansion are used, while in the eikonal limit $\ell\gg1$ the approximation becomes asymptotically exact. Here we used the 6th WKB order, as it usually produces sufficient accuracy when compared with more accurate methods.

\section{Grey-body factors and correspondence with quasinormal modes}

In the eikonal regime, the scattering problem admits a simple analytic description, revealing a direct relation between the transmission probability and the black hole's fundamental quasinormal mode. In this limit, the grey-body factor can be expressed as \cite{Konoplya:2024lir,Konoplya:2024vuj}
\begin{eqnarray}\label{transmission-eikonal}
\Gamma_{\ell}(\Omega)\equiv |T|^2 &=&
\left(1+e^{2\pi\dfrac{\Omega^2-\re{\omega_0}^2}{4\re{\omega_0}\im{\omega_0}}}\right)^{-1}
+ \Order{\ell^{-1}} .
\end{eqnarray}

This relation exemplifies the well–known correspondence between grey–body factors and quasinormal modes in the eikonal limit. For large multipole number $\ell$, both quantities are determined by the same characteristics of the effective potential near its maximum. Specifically, the real part of the fundamental quasinormal frequency $\re{\omega_0}$ is associated with the oscillation frequency of waves trapped near the peak of the potential barrier, while the imaginary part $\im{\omega_0}$ characterizes the instability timescale. This correspondence has been extensively tested in numerous recent studies \cite{Han:2025cal,Lutfuoglu:2025blw,Malik:2025dxn,Dubinsky:2024vbn,Bolokhov:2024otn,Skvortsova:2024msa,Malik:2024cgb,Lutfuoglu:2025hjy,Bolokhov:2025lnt,Lutfuoglu:2025ohb,Lutfuoglu:2025ldc,Dubinsky:2025nxv,Malik:2025erb}, typically showing excellent accuracy for higher multipole numbers.

Our computed grey-body factors (see figs. 1-4) show that the considered regular black-hole metrics yield only very minor deviations from the Schwarzschild case. This behavior can be understood from the structure of the effective potential governing scalar-field propagation. In both models, the metric differs from the Schwarzschild geometry mainly in the near-horizon region, while the position and overall shape of the potential barrier remain almost unchanged. Since the transmission probability is primarily determined by the properties of the potential near its maximum, such localized modifications produce only minor corrections to the grey-body factors. Consequently, both the fundamental quasinormal mode and the corresponding transmission spectra remain very close to their Schwarzschild counterparts. Moreover, the correspondence between grey-body factors and quasinormal modes works well even for $\ell=1$ and becomes essentially exact for $\ell=2$ (see figs. 2 and 4), with the difference between the WKB approximation and the quasinormal-mode prediction becomes negligible. However, for the monopole case $\ell=0$ the WKB approximation is known to be inaccurate, rendering the correspondence unreliable in this regime. We therefore restrict our analysis to $\ell \ge 1$. Overall, our results indicate that grey-body factors are remarkably stable \cite{Oshita:2024fzf,Oshita:2023cjz,Rosato:2024arw,Konoplya:2025ixm} under regularization of the geometry. This stands in contrast to the higher overtones of quasinormal modes, which are known to be far more sensitive to near-horizon modifications of the spacetime \cite{Konoplya:2022pbc}.

\begin{acknowledgments}
The author acknowledges the University of Seville for their support through the Plan-US of aid to Ukraine.
\end{acknowledgments}

\bibliography{bibliography}

@article{Konoplya:2022pbc,
    author = "Konoplya, R. A. and Zhidenko, A.",
    title = "{First few overtones probe the event horizon geometry}",
    eprint = "2209.00679",
    archivePrefix = "arXiv",
    primaryClass = "gr-qc",
    doi = "10.1016/j.jheap.2024.10.015",
    journal = "JHEAp",
    volume = "44",
    pages = "419--426",
    year = "2024"
}

@article{Iyer:1986np,
    author = "Iyer, Sai and Will, Clifford M.",
    title = "{Black Hole Normal Modes: A {WKB} Approach. 1. Foundations and Application of a Higher Order {WKB} Analysis of Potential Barrier Scattering}",
    reportNumber = "Print-86-1482 (WASH. U., ST. LOUIS)",
    doi = "10.1103/PhysRevD.35.3621",
    journal = "Phys. Rev. D",
    volume = "35",
    pages = "3621",
    year = "1987"
}

@article{Konoplya:2003ii,
    author = "Konoplya, R. A.",
    title = "{Quasinormal behavior of the d-dimensional Schwarzschild black hole and higher order WKB approach}",
    eprint = "gr-qc/0303052",
    archivePrefix = "arXiv",
    doi = "10.1103/PhysRevD.68.024018",
    journal = "Phys. Rev. D",
    volume = "68",
    pages = "024018",
    year = "2003"
}

@article{Matyjasek:2017psv,
    author = "Matyjasek, Jerzy and Opala, Michał",
    title = "{Quasinormal modes of black holes. The improved semianalytic approach}",
    eprint = "1704.00361",
    archivePrefix = "arXiv",
    primaryClass = "gr-qc",
    doi = "10.1103/PhysRevD.96.024011",
    journal = "Phys. Rev. D",
    volume = "96",
    number = "2",
    pages = "024011",
    year = "2017"
}

@article{Hayward:2005gi,
    author = "Hayward, Sean A.",
    title = "{Formation and evaporation of regular black holes}",
    eprint = "gr-qc/0506126",
    archivePrefix = "arXiv",
    doi = "10.1103/PhysRevLett.96.031103",
    journal = "Phys. Rev. Lett.",
    volume = "96",
    pages = "031103",
    year = "2006"
}

@inproceedings{Ansoldi:2008jw,
    author = "Ansoldi, Stefano",
    title = "{Spherical black holes with regular center: A Review of existing models including a recent realization with Gaussian sources}",
    booktitle = "{Conf. on BHs and Naked Singularities}",
    eprint = "0802.0330",
    archivePrefix = "arXiv",
    primaryClass = "gr-qc",
    reportNumber = "KUNS-2108",
    month = "2",
    year = "2008"
}

@article{Dymnikova:1992ux,
    author = "Dymnikova, I.",
    title = "{Vacuum nonsingular black hole}",
    doi = "10.1007/BF00760226",
    journal = "Gen. Rel. Grav.",
    volume = "24",
    pages = "235--242",
    year = "1992"
}

@article{Bronnikov:2000vy,
    author = "Bronnikov, Kirill A.",
    title = "{Regular magnetic black holes and monopoles from nonlinear electrodynamics}",
    eprint = "gr-qc/0006014",
    archivePrefix = "arXiv",
    doi = "10.1103/PhysRevD.63.044005",
    journal = "Phys. Rev. D",
    volume = "63",
    pages = "044005",
    year = "2001"
}

@article{Bronnikov:2024izh,
    author = "Bronnikov, Kirill A.",
    title = "{Regular black holes as an alternative to black bounce}",
    eprint = "2404.14816",
    archivePrefix = "arXiv",
    primaryClass = "gr-qc",
    doi = "10.1103/PhysRevD.110.024021",
    journal = "Phys. Rev. D",
    volume = "110",
    number = "2",
    pages = "024021",
    year = "2024"
}

@article{AyonBeato:1998ub,
    author = "Ayon-Beato, Eloy and Garcia, Alberto",
    title = "{Regular black hole in general relativity coupled to nonlinear electrodynamics}",
    eprint = "gr-qc/9911046",
    archivePrefix = "arXiv",
    doi = "10.1103/PhysRevLett.80.5056",
    journal = "Phys. Rev. Lett.",
    volume = "80",
    pages = "5056--5059",
    year = "1998"
}

@inproceedings{Bardeen:1968,
    author = "Bardeen, J. M.",
    title = "{Non-singular general-relativistic gravitational collapse}",
    booktitle = "{Proc. Int. Conf. GR5}",
    year = "1968",
    address = "Tbilisi, USSR"
}

@article{Bronnikov:2005gm,
    author = "Bronnikov, K. A. and Fabris, J. C.",
    title = "{Regular phantom black holes}",
    eprint = "gr-qc/0511109",
    archivePrefix = "arXiv",
    doi = "10.1103/PhysRevLett.96.251101",
    journal = "Phys. Rev. Lett.",
    volume = "96",
    pages = "251101",
    year = "2006"
}

@article{Malik:2025dxn,
    author = "Malik, Zainab",
    title = "{Gravitational Perturbations of the Hayward Spacetime and Testing the Correspondence between Quasinormal Modes and Grey-body Factors}",
    eprint = "2508.19178",
    archivePrefix = "arXiv",
    primaryClass = "gr-qc",
    doi = "10.1007/s10773-025-06198-w",
    journal = "Int. J. Theor. Phys.",
    volume = "64",
    number = "11",
    pages = "314",
    year = "2025"
}

@article{Schutz:1985km,
    author = "Schutz, Bernard F. and Will, Clifford M.",
    title = "{BLACK HOLE NORMAL MODES: A SEMIANALYTIC APPROACH}",
    reportNumber = "PRINT-85-0063 (WASH.U.,ST.LOUIS)",
    doi = "10.1086/184453",
    journal = "Astrophys. J. Lett.",
    volume = "291",
    pages = "L33--L36",
    year = "1985"
}

@article{Bueno:2025tli,
    author = "Bueno, Pablo and Hennigar, Robie A. and Murcia, {\'A}ngel J. and Vicente-Cano, Aitor",
    title = "{Buchdahl limits in theories with regular black holes}",
    eprint = "2512.19796",
    archivePrefix = "arXiv",
    primaryClass = "gr-qc",
    month = "12",
    year = "2025"
}

@article{Bueno:2024eig,
    author = "Bueno, Pablo and Cano, Pablo A. and Hennigar, Robie A. and Murcia, {\'A}ngel J.",
    title = "{Dynamical Formation of Regular Black Holes}",
    eprint = "2412.02742",
    archivePrefix = "arXiv",
    primaryClass = "gr-qc",
    doi = "10.1103/PhysRevLett.134.181401",
    journal = "Phys. Rev. Lett.",
    volume = "134",
    number = "18",
    pages = "181401",
    year = "2025"
}

@article{Bueno:2025zaj,
    author = "Bueno, Pablo and Cano, Pablo A. and Hennigar, Robie A. and Murcia, {\'A}ngel J.",
    title = "{Regular black hole formation in four-dimensional nonpolynomial gravities}",
    eprint = "2509.19016",
    archivePrefix = "arXiv",
    primaryClass = "gr-qc",
    doi = "10.1103/8f3j-zcxh",
    journal = "Phys. Rev. D",
    volume = "113",
    number = "2",
    pages = "024019",
    year = "2026"
}

@article{Bueno:2024dgm,
    author = "Bueno, Pablo and Cano, Pablo A. and Hennigar, Robie A.",
    title = "{Regular black holes from pure gravity}",
    eprint = "2403.04827",
    archivePrefix = "arXiv",
    primaryClass = "gr-qc",
    doi = "10.1016/j.physletb.2025.139260",
    journal = "Phys. Lett. B",
    volume = "861",
    pages = "139260",
    year = "2025"
}

@article{Borissova:2026wmn,
    author = "Borissova, Johanna and Carballo-Rubio, Ra{\'u}l",
    title = "{Regular black holes from pure gravity in four dimensions}",
    eprint = "2602.16773",
    archivePrefix = "arXiv",
    primaryClass = "gr-qc",
    reportNumber = "Imperial/TP/2026/JB/02",
    month = "2",
    year = "2026"
}

@article{Lutfuoglu:2025ohb,
    author = {L{\"u}tf{\"u}o{\u{g}}lu, B. C.},
    title = "{Quasinormal modes and gray-body factors for gravitational perturbations in asymptotically safe gravity}",
    eprint = "2505.06966",
    archivePrefix = "arXiv",
    primaryClass = "gr-qc",
    doi = "10.1140/epjc/s10052-026-15290-2",
    journal = "Eur. Phys. J. C",
    volume = "86",
    number = "1",
    pages = "39",
    year = "2026"
}

@article{Dymnikova:2015yma,
    author = "Dymnikova, Irina and Khlopov, Maxim",
    title = "{Regular black hole remnants and graviatoms with de Sitter interior as heavy dark matter candidates probing inhomogeneity of early universe}",
    eprint = "1510.01351",
    archivePrefix = "arXiv",
    primaryClass = "gr-qc",
    doi = "10.1142/S0218271815450029",
    journal = "Int. J. Mod. Phys. D",
    volume = "24",
    number = "13",
    pages = "1545002",
    year = "2015"
}

@article{Spina:2025wxb,
    author = "Spina, Andrea",
    title = "{Black Holes in Asymptotic Safety: A Review of Solutions and Phenomenology}",
    eprint = "2510.14552",
    archivePrefix = "arXiv",
    primaryClass = "gr-qc",
    doi = "10.53941/ijgtp.2025.100008",
    journal = "Int. J. Grav. Theor. Phys.",
    volume = "1",
    number = "1",
    pages = "8",
    year = "2025"
}

@article{Konoplya:2025ect,
    author = "Konoplya, R. A. and Zhidenko, A.",
    title = "{Dark matter halo as a source of regular black-hole geometries}",
    eprint = "2511.03066",
    archivePrefix = "arXiv",
    primaryClass = "gr-qc",
    doi = "10.1103/7ptp-9j1t",
    journal = "Phys. Rev. D",
    volume = "113",
    number = "4",
    pages = "043011",
    year = "2026"
}

@article{Arbelaez:2026eaz,
    author = "Arbelaez, Juan Pablo",
    title = "{Grey-body factors of higher dimensional regular black holes in quasi-topological theories}",
    eprint = "2601.22340",
    archivePrefix = "arXiv",
    primaryClass = "gr-qc",
    month = "1",
    year = "2026"
}

@article{Konoplya:2024lir,
    author = "Konoplya, R. A. and Zhidenko, A.",
    title = "{Correspondence between grey-body factors and quasinormal modes}",
    eprint = "2406.11694",
    archivePrefix = "arXiv",
    primaryClass = "gr-qc",
    doi = "10.1088/1475-7516/2024/09/068",
    journal = "JCAP",
    volume = "09",
    pages = "068",
    year = "2024"
}

@article{Konoplya:2024vuj,
    author = "Konoplya, R. A. and Zhidenko, A.",
    title = "{Correspondence between grey-body factors and quasinormal frequencies for rotating black holes}",
    eprint = "2408.11162",
    archivePrefix = "arXiv",
    primaryClass = "gr-qc",
    doi = "10.1016/j.physletb.2025.139288",
    journal = "Phys. Lett. B",
    volume = "861",
    pages = "139288",
    year = "2025"
}

@article{Bolokhov:2024otn,
    author = "Bolokhov, S. V. and Skvortsova, Milena",
    title = "{Correspondence between quasinormal modes and grey-body factors of spherically symmetric traversable wormholes}",
    eprint = "2412.11166",
    archivePrefix = "arXiv",
    primaryClass = "gr-qc",
    doi = "10.1088/1475-7516/2025/04/025",
    journal = "JCAP",
    volume = "04",
    pages = "025",
    year = "2025"
}

@article{Malik:2024cgb,
    author = "Malik, Zainab",
    title = "{Correspondence between quasinormal modes and grey-body factors for massive fields in Schwarzschild-de~Sitter spacetime}",
    eprint = "2412.19443",
    archivePrefix = "arXiv",
    primaryClass = "gr-qc",
    doi = "10.1088/1475-7516/2025/04/042",
    journal = "JCAP",
    volume = "04",
    pages = "042",
    year = "2025"
}

@article{Oshita:2023cjz,
    author = "Oshita, Naritaka",
    title = "{Greybody factors imprinted on black hole ringdowns: An alternative to superposed quasinormal modes}",
    eprint = "2309.05725",
    archivePrefix = "arXiv",
    primaryClass = "gr-qc",
    reportNumber = "YITP-23-111, RIKEN-iTHEMS-Report-23",
    doi = "10.1103/PhysRevD.109.104028",
    journal = "Phys. Rev. D",
    volume = "109",
    number = "10",
    pages = "104028",
    year = "2024"
}

@article{Rosato:2024arw,
    author = "Rosato, Romeo Felice and Destounis, Kyriakos and Pani, Paolo",
    title = "{Ringdown stability: Graybody factors as stable gravitational-wave observables}",
    eprint = "2406.01692",
    archivePrefix = "arXiv",
    primaryClass = "gr-qc",
    doi = "10.1103/PhysRevD.110.L121501",
    journal = "Phys. Rev. D",
    volume = "110",
    number = "12",
    pages = "L121501",
    year = "2024"
}

@article{Oshita:2024fzf,
    author = "Oshita, Naritaka and Takahashi, Kazufumi and Mukohyama, Shinji",
    title = "{Stability and instability of the black hole greybody factors and ringdowns against a small-bump correction}",
    eprint = "2406.04525",
    archivePrefix = "arXiv",
    primaryClass = "gr-qc",
    reportNumber = "YITP-24-69, IPMU24-0025, RIKEN-iTHEMS-Report-24",
    doi = "10.1103/PhysRevD.110.084070",
    journal = "Phys. Rev. D",
    volume = "110",
    number = "8",
    pages = "084070",
    year = "2024"
}

@article{Konoplya:2025ixm,
    author = "Konoplya, Roman A. and Pappas, Thomas D.",
    title = "{Dirty black holes, clean signals: near-horizon vs.~environmental effects on grey-body factors and Hawking radiation}",
    eprint = "2507.01954",
    archivePrefix = "arXiv",
    primaryClass = "gr-qc",
    doi = "10.1088/1475-7516/2026/02/038",
    journal = "JCAP",
    volume = "02",
    pages = "038",
    year = "2026"
}

@article{Page:1976df,
    author = "Page, Don N.",
    title = "{Particle Emission Rates from a Black Hole: Massless Particles from an Uncharged, Nonrotating Hole}",
    doi = "10.1103/PhysRevD.13.198",
    journal = "Phys. Rev. D",
    volume = "13",
    pages = "198--206",
    year = "1976"
}

@article{Page:1976ki,
    author = "Page, Don N.",
    title = "{Particle Emission Rates from a Black Hole. 2. Massless Particles from a Rotating Hole}",
    doi = "10.1103/PhysRevD.14.3260",
    journal = "Phys. Rev. D",
    volume = "14",
    pages = "3260--3273",
    year = "1976"
}

@article{Tsuda:2026xjc,
    author = "Tsuda, Ren and Suzuki, Ryotaku and Tomizawa, Shinya",
    title = "{Fan-Wang type regular black holes in Quasi-Topological Gravity}",
    eprint = "2602.16754",
    archivePrefix = "arXiv",
    primaryClass = "gr-qc",
    reportNumber = "TTI-MATHPHYS-39",
    month = "2",
    year = "2026"
}

@article{Bonanno:2000ep,
    author = "Bonanno, Alfio and Reuter, Martin",
    title = "{Renormalization group improved black hole space-times}",
    eprint = "hep-th/0002196",
    archivePrefix = "arXiv",
    reportNumber = "INFN-CT-03-00, MZ-TH-00-04",
    doi = "10.1103/PhysRevD.62.043008",
    journal = "Phys. Rev. D",
    volume = "62",
    pages = "043008",
    year = "2000"
}

@article{Konoplya:2026gim,
    author = "Konoplya, R. A.",
    title = "{Quasinormal modes of four-dimensional regular black holes in quasi-topological gravity: Overtones' outburst via WKB method}",
    eprint = "2603.03189",
    archivePrefix = "arXiv",
    primaryClass = "gr-qc",
    month = "3",
    year = "2026"
}

@article{Cvetic:1997xv,
    author = "Cvetic, Mirjam and Larsen, Finn",
    title = "{Grey body factors for rotating black holes in four-dimensions}",
    eprint = "hep-th/9706071",
    archivePrefix = "arXiv",
    reportNumber = "UPR-0755-T",
    doi = "10.1016/S0550-3213(97)00541-5",
    journal = "Nucl. Phys. B",
    volume = "506",
    pages = "107--120",
    year = "1997"
}

@article{Gubser:1996zp,
    author = "Gubser, Steven S. and Klebanov, Igor R.",
    title = "{Four-dimensional grey body factors and the effective string}",
    eprint = "hep-th/9609076",
    archivePrefix = "arXiv",
    reportNumber = "PUPT-1648",
    doi = "10.1103/PhysRevLett.77.4491",
    journal = "Phys. Rev. Lett.",
    volume = "77",
    pages = "4491--4494",
    year = "1996"
}

@article{Klebanov:1997cx,
    author = "Klebanov, Igor R. and Mathur, Samir D.",
    title = "{Black hole grey body factors and absorption of scalars by effective strings}",
    eprint = "hep-th/9701187",
    archivePrefix = "arXiv",
    reportNumber = "PUPT-1679",
    doi = "10.1016/S0550-3213(97)00287-3",
    journal = "Nucl. Phys. B",
    volume = "500",
    pages = "115--132",
    year = "1997"
}

@article{Kanti:2014vsa,
    author = "Kanti, Panagiota and Winstanley, Elizabeth",
    title = "{Hawking Radiation from Higher-dimensional Black Holes}",
    eprint = "1402.3952",
    archivePrefix = "arXiv",
    primaryClass = "hep-th",
    doi = "10.1007/978-3-319-10852-0_8",
    journal = "Fundam. Theor. Phys.",
    volume = "178",
    pages = "229--265",
    year = "2015"
}

@article{Malik:2025erb,
    author = "Malik, Zainab",
    title = "{Grey-Body Factors for Scalar and Dirac Fields in the Euler-Heisenberg Electrodynamics}",
    eprint = "2509.15995",
    archivePrefix = "arXiv",
    primaryClass = "gr-qc",
    doi = "10.53941/ijgtp.2025.100006",
    journal = "Int. J. Grav. Theor. Phys.",
    volume = "1",
    number = "1",
    pages = "6",
    year = "2025"
}

@article{Dubinsky:2025nxv,
    author = "Dubinsky, Alexey",
    title = "{Gravitational perturbations of Dymnikova black holes: Grey-body factors and absorption cross-sections}",
    eprint = "2509.11017",
    archivePrefix = "arXiv",
    primaryClass = "gr-qc",
    doi = "10.1016/j.aop.2025.170299",
    journal = "Annals Phys.",
    volume = "485",
    pages = "170299",
    year = "2026"
}

@article{Dubinsky:2024vbn,
    author = "Dubinsky, Alexey",
    title = "{Gray-body factors for gravitational and electromagnetic perturbations around Gibbons{\textendash}Maeda{\textendash}Garfinkle{\textendash}Horowitz{\textendash}Strominger black holes}",
    eprint = "2412.00625",
    archivePrefix = "arXiv",
    primaryClass = "gr-qc",
    doi = "10.1142/S0217732325501111",
    journal = "Mod. Phys. Lett. A",
    volume = "40",
    number = "28",
    pages = "2550111",
    year = "2025"
}

@article{Lutfuoglu:2025blw,
    author = {L{\"u}tf{\"u}o{\u{g}}lu, Bekir Can and Saka, Erdin{\c{c}} Ula{\c{s}} and Shermatov, Abubakir and Rayimbaev, Javlon and Ibragimov, Inomjon and Muminov, Sokhibjan},
    title = "{Proper-time approach in asymptotic safety via black hole quasinormal modes and grey-body factors}",
    eprint = "2509.15923",
    archivePrefix = "arXiv",
    primaryClass = "gr-qc",
    doi = "10.1140/epjc/s10052-025-14950-z",
    journal = "Eur. Phys. J. C",
    volume = "85",
    number = "10",
    pages = "1190",
    year = "2025"
}

@article{Lutfuoglu:2025ldc,
    author = {L{\"u}tf{\"u}o{\u{g}}lu, Bekir Can},
    title = "{Black Holes in Proca-Gauss-Bonnet Gravity with Primary Hair: Particle Motion, Shadows, and Grey-Body Factors}",
    eprint = "2507.09246",
    archivePrefix = "arXiv",
    primaryClass = "gr-qc",
    doi = "10.53941/ijgtp.2025.100004",
    journal = "Int. J. Grav. Theor. Phys.",
    volume = "1",
    number = "1",
    pages = "4",
    year = "2025"
}

@article{Lutfuoglu:2025ljm,
    author = {L{\"u}tf{\"u}o{\u{g}}lu, B. C.},
    title = "{Non-minimal Einstein{\textendash}Yang{\textendash}Mills black holes: fundamental quasinormal mode and grey-body factors versus outburst of overtones}",
    eprint = "2504.18482",
    archivePrefix = "arXiv",
    primaryClass = "gr-qc",
    doi = "10.1140/epjc/s10052-025-14380-x",
    journal = "Eur. Phys. J. C",
    volume = "85",
    number = "6",
    pages = "630",
    year = "2025"
}

@article{Lutfuoglu:2025hjy,
    author = {L{\"u}tf{\"u}o{\u{g}}lu, B. C.},
    title = "{Long-lived quasinormal modes and gray-body factors of black holes and wormholes in dark matter inspired Weyl gravity}",
    eprint = "2503.16087",
    archivePrefix = "arXiv",
    primaryClass = "gr-qc",
    doi = "10.1140/epjc/s10052-025-14210-0",
    journal = "Eur. Phys. J. C",
    volume = "85",
    number = "5",
    pages = "486",
    year = "2025"
}

@article{Bolokhov:2025lnt,
    author = "Bolokhov, S. V. and Skvortsova, Milena",
    title = "{Gravitational Quasinormal Modes and Grey-Body Factors of Bonanno{\textendash}Reuter Regular Black Holes}",
    eprint = "2507.07196",
    archivePrefix = "arXiv",
    primaryClass = "gr-qc",
    doi = "10.53941/ijgtp.2025.100003",
    journal = "Int. J. Grav. Theor. Phys.",
    volume = "1",
    number = "1",
    pages = "3",
    year = "2025"
}

@article{Skvortsova:2024msa,
    author = "Skvortsova, Milena",
    title = "{Quantum corrected black holes: testing the correspondence between grey-body factors and quasinormal modes}",
    eprint = "2411.06007",
    archivePrefix = "arXiv",
    primaryClass = "gr-qc",
    doi = "10.1140/epjc/s10052-025-14589-w",
    journal = "Eur. Phys. J. C",
    volume = "85",
    number = "8",
    pages = "854",
    year = "2025"
}

@article{Kanti:2017ubd,
    author = "Kanti, Panagiota and Pappas, Thomas",
    title = "{Effective temperatures and radiation spectra for a higher-dimensional Schwarzschild{\textendash}de Sitter black hole}",
    eprint = "1705.09108",
    archivePrefix = "arXiv",
    primaryClass = "hep-th",
    doi = "10.1103/PhysRevD.96.024038",
    journal = "Phys. Rev. D",
    volume = "96",
    number = "2",
    pages = "024038",
    year = "2017"
}

@article{Han:2025cal,
    author = "Han, Hyewon and Gwak, Bogeun",
    title = "{Correspondence between quasinormal modes and grey-body factors in five-dimensional black holes}",
    eprint = "2508.12989",
    archivePrefix = "arXiv",
    primaryClass = "gr-qc",
    month = "8",
    year = "2025"
}
\end{document}